\documentclass[prb,a4paper,twocolumn,showpacs,floatfix,superscriptaddress,amsmath,amsfonts,amssymb,preprintnumbers,citeautoscript,aps,longbibliography]{revtex4-2}
\pdfoutput=1

\usepackage{hyperref}
\hypersetup{backref=true,pagebackref=true,hyperindex=true,colorlinks=true,breaklinks=true,
urlcolor=blue,linkcolor=blue,bookmarks=true,bookmarksopen=true,citecolor=red}

\usepackage[T1]{fontenc}\usepackage[latin1]{inputenc}

\usepackage{amsmath}
\usepackage{amssymb}
\usepackage{bbold} 
\usepackage{graphicx}
\usepackage{dcolumn,graphicx,color,booktabs,microtype,afterpage}

\usepackage{bm}
\usepackage{braket}

\usepackage{times}

\usepackage{color}
\usepackage[normalem]{ulem}
\begin{document}

\title{ Detection and control  of electronic orbital magnetism by spin waves in honeycomb ferromagnets}

\author{Lichuan Zhang}
\email[Corresponding author:~]{lichuan.zhang@ujs.edu.cn}
\affiliation{School of Physics and Electronic Engineering, Jiangsu University, Zhenjiang 212013, China}
\affiliation{Jiangsu Engineering Research Center on Quantum Perception and Intelligent
Detection of Agricultural Information, Zhenjiang 212013, China}

\author{Lu Zhang}
\affiliation{School of Physics and Electronic Engineering, Jiangsu University, Zhenjiang 212013, China}
\affiliation{Jiangsu Engineering Research Center on Quantum Perception and Intelligent
Detection of Agricultural Information, Zhenjiang 212013, China}

\author{Dongwook Go}
\affiliation{Peter Gr\"unberg Institut and Institute for Advanced Simulation, Forschungszentrum J\"ulich and JARA, 52425 J\"ulich, Germany}
\affiliation{Institute of Physics, Johannes Gutenberg University Mainz, 55099 Mainz, Germany}

\author{Chengwang Niu}
\affiliation{School of Physics, State Key Laboratory of Crystal Materials, Shandong University, Jinan 250100, China}

\author{Wulf Wulfhekel}
\affiliation{Physikalisches Institut, Karlsruhe Institute of Technology, 76131 Karlsruhe, Germany}

\author{Peng Li}
\affiliation{School of Microelectronics,
University of Science and Technology of China Hefei 230026, China}

\author{Yuanping Chen}\email[Corresponding author:~]{chenyp@ujs.edu.cn}
\affiliation{School of Physics and Electronic Engineering, Jiangsu University, Zhenjiang 212013, China}
\affiliation{Jiangsu Engineering Research Center on Quantum Perception and Intelligent
Detection of Agricultural Information, Zhenjiang 212013, China}
\author{Yuriy Mokrousov}
\email[Corresponding author:~]{y.mokrousov@fz-juelich.de}
\affiliation{Peter Gr\"unberg Institut and Institute for Advanced Simulation, Forschungszentrum J\"ulich and JARA, 52425 J\"ulich, Germany}
\affiliation{Institute of Physics, Johannes Gutenberg University Mainz, 55099 Mainz, Germany}

\author{Lizhi Zhang}\email[Corresponding author:~]{zhanglz@nanoctr.cn}
\affiliation{Laboratory of Theoretical and Computational Nanoscience, National Center for Nanoscience and Technology, Beijing, 100190, China}

\begin{abstract}
\noindent 

Exploring and manipulating the orbital degrees of freedom in solids has become a fascinating research topic in modern magnetism. Here, we demonstrate that spin waves can provide a way to control electronic orbital magnetism by the mechanism of scalar spin chirality, allowing for experimental detection using techniques such as the magneto-optical Kerr effect and scanning transmission electron microscopy. By applying linear spin wave theory, we uncover that electronic magnon-driven orbital magnetization is extremely sensitive to the character of the magnonic excitations. Furthermore, we show that both the induced electronic orbital magnetism and the Nernst transport properties of the orbital angular momentum can be regulated by the strength of the Dzyaloshinskii-Moriya interaction, Kitaev interaction, as well as the direction and magnitude of the external magnetic field. We argue that magnon-mediated electronic orbital magnetism presents an emergent variable which has to be taken into account when considering the physics of coupling magnonic excitiations to phonons and light.
\end{abstract}

\maketitle


Recently,  orbitronics~\cite{orbitronicsGo} as a field started to attract significant attention, owing to strong theoretical and experimental interest in orbital currents~\cite{orbitaltorque} and related phenomena such as  orbital torques~\cite{yang2024orbital}, the
orbital Hall effect~\cite{OHE2023,Go2018Intrinsic}, the orbital Rashba effect~\cite{2021Nontrivial,Adamantopoulos2024} and orbital magnetoresistance~\cite{Ding2022}. While in the ground state of collinear magnets the orbital magnetization arises as a consequence of spin-orbit coupling (SOC), it has been realized that in non-collinear magnets the unquenched orbital angular momentum can emerge even without SOC from nonvanishing scalar spin chirality (SSC), defined as $\chi=\mathbf S_1 \cdot(\mathbf S_2 \times\mathbf S_3 )$, where three neighboring spins form a triangle~\cite{Taguchi2001Spin,2001Orbital}. The SSC acts like a magnetic field, influencing electron dynamics similarly to a spin-dependent magnetic field ~\cite{nagaosa2013topological}, thus providing an alternative mechanism for orbital magnetism that does not rely on SOC. 
The emergence of scalar-chirality-driven orbital magnetization, commonly referred to as topological orbital moment (TOM), has been explored in various spin systems~\cite{hoffmann2015topological,hanke2016role,hanke2017prototypical,lux2018engineering,grytsiuk2020topological,PhysRevB.108.L180411,PhysRevB.92.020401,nickel2024antiferromagnetic}.  
As the SSC is inherent to
skyrmions~\cite{lux2018engineering, Matthias19} and diverse types of frustrated magnets~\cite{Taguchi2001Spin, Satoshi09} $-$ playing a crucial rule in understanding phenomena such as the topological Hall effect~\cite{THE09},  topological magneto-optical effect~\cite{wanxiang_NC}, and the quantum topological Hall effects~\cite{qthe} $-$ TOM emerges as a very promising platform for the design of orbital magnetism in non-collinear magnets.

In the ground state of collinear magnets, TOM is forbidden by symmetry. However, in analogy to the logic of orbitronics which relies on the fact that orbital magnetism can be promoted when out of equilibrium, non-relativistic electronic orbital magnetism may arise when collinear magnets are subject to thermal fluctuations. In particular, we
recently have shown that electronic orbital magnetism can be imprinted and driven by the magnonic excitations of kagome ferromagnets (FMs)~\cite{zhang_imprinting_2020}. Recent evidence of magnon-mediated orbital magnetism has been obtained in the quasi-two-dimensional topological magnon insulator Cu(1,3-bdc)~\cite{chisnell2016magnetic,lipeng-2022}. 
However, the experimental detection of magnon-related TOM and control of its properties on different spin lattices remains elusive. 
Besides, while spin waves (SW) can be generated and injected into magnets, influencing their dynamic spin structure~\cite{pirro2021advances},
the nature of the relationship between TOM and various SW modes is largely unknown. Therefore, it is essential to investigate the characteristics of TOM imprinted by different SW modes on diverse spin lattices, alongside the effects of external magnetic field or temperature modulation.

The physics of spin excitations in magnetic two-dimensional (2D) materials is very rich, with magnons observed in systems such as 2D honeycomb magnets~\cite{cenker2021direct}. Ferromagnetic honeycomb materials, e.g., CrI$_3$~\cite{2017Layer} and 
CrGeTe$_3$~\cite{CrGeTe32017}, exhibit a wide range of spin interactions, such as the Heisenberg exchange interaction, Dzyaloshinskii-Moriya interaction (DMI)~\cite{2021Topological}, and Kitaev interaction~\cite{lee2020fundamental}. Additionally, numerous frustrated spin textures with significant SSC, such as skyrmions~\cite{PhysRevB.98.180407} and multiple-$q$ states~\cite{PhysRevB.100.224404}, have been reported and observed on a honeycomb lattice. These diverse magnetic ground states~\cite{PhysRevLett.117.277202} and topological magnonic phases~\cite{2021Topological, zhang2021kitaev, bai2024coupled} demonstrate that the honeycomb lattice is an excellent platform for studying the relationship between TOM, SW, and magnonic topology. 

\begin{figure}
\includegraphics[width=0.95\linewidth]{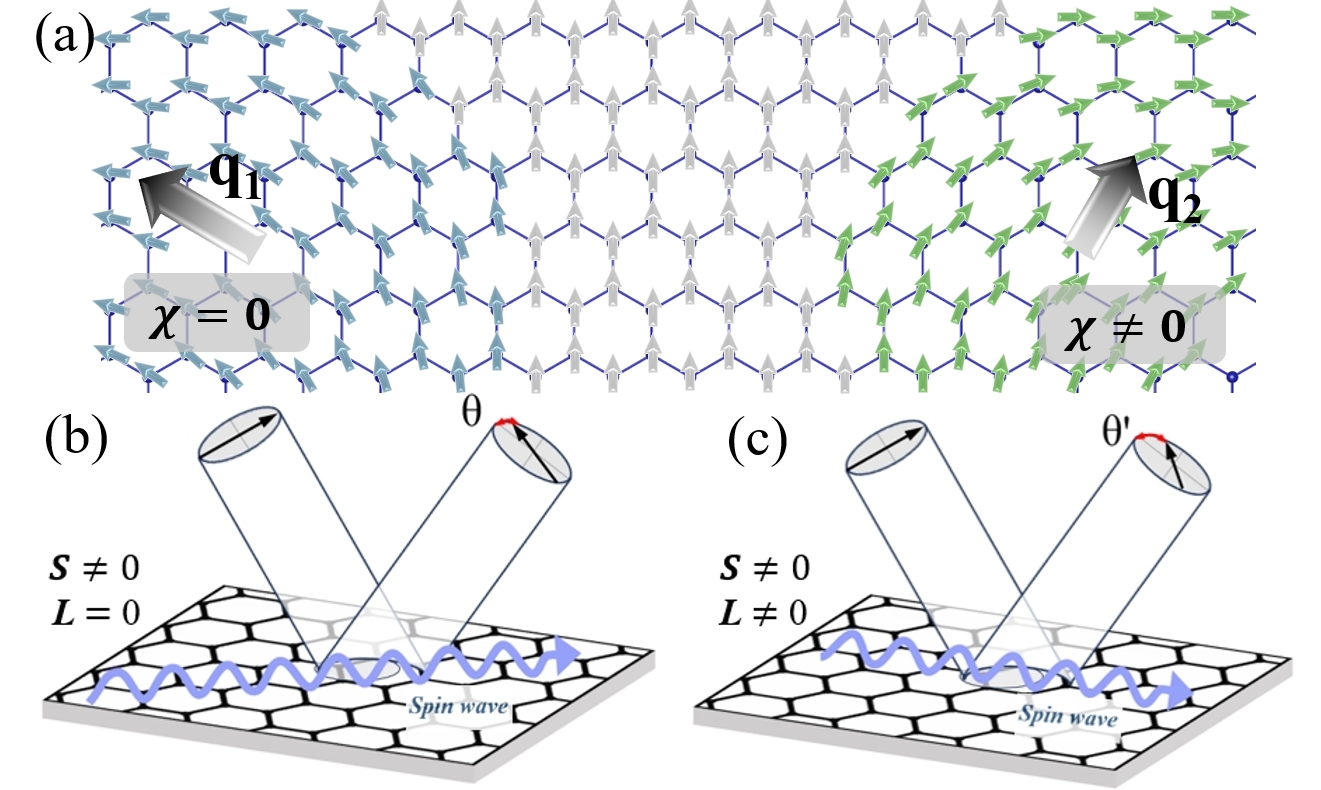}
    \caption{(a)~Imprinting electronic orbital magnetism by magnons in a honeycomb ferromagnetic lattice with different $q$-vectors. The scalar spin chirality for $\mathbf q_1$ is $0$~(SW along armchair direction), and the scalar spin chirality for $\mathbf q_2$ is nonzero (b). (b, c)~ The diagram of detecting orbital moment modulated by SW via magneto-optical Kerr effect. If a SW is injected or generated in a 2D honeycomb ferromagnet, the generated orbital angular moment can be controlled when the propagation direction of the SW is changed. } 
\label{fig1}
\end{figure} 

\begin{figure*}
\centering\vspace{-1pt}   \includegraphics[width=0.95\linewidth]{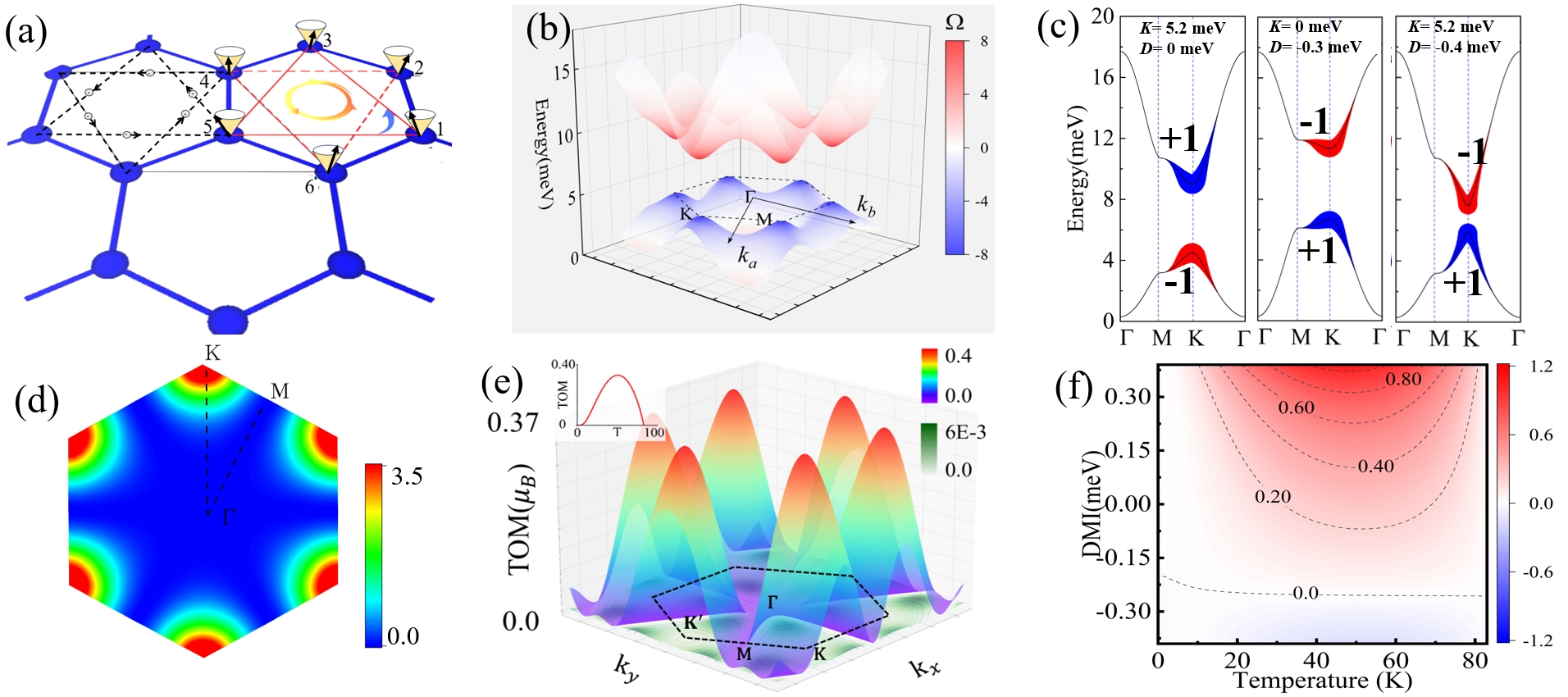}
    \caption{ (a) The schematic structure of the honeycomb lattice in an oblique view. 
    The arrows along the black dots indicate the orientations of the second-nearest-neighbor bonds, with the DMI vector along the out-of-plane direction. Two types of SSC are classified by different colors.  (b) The magnonic dispersion and Berry curvature distribution 
    with Heishenberg-Kitaev model. 
    The color map refers to the value of Berry curvature ranging from $-8$ to 8 arb. units. 
    (c) Fat band analysis for the magnonic bands, with the values of the DMI and Kitaev interactions specified at the top. Red and blue colors separately represent positive and negative signs of $L_{n\mathbf k}^\mathrm{TOM}$, and the line thickness denotes the corresponding magnitude. Bold integers indicate Chern numbers. 
    (d)~The distribution of $L_{1\mathbf k}^\mathrm{TOM}$  in the first Brillouin zone (BZ).
    (e)~The distribution of $\sum_n\ell_n(\mathbf k, T)$ at 5~K (green color) and 20~K (rainbow color), separately.   The insert shows the overall TOM $\braket{L^\mathrm{TOM}}_\mathrm T$ as a function of temperature. (f)~$\braket{L^\mathrm{TOM}}_\mathrm T$ as a function of  $D$ and temperature. For (b, d-f), parameters are chosen as  $K= 5.2$ meV, $J=0.2$ meV, and the unit of the color map for (d-f) is $\mu_\mathbf{B}$. }
    \label{fig2}
\end{figure*}

In this work, we systematically investigate the magnitude and transport properties of TOM modulated by magnons on a ferromagnetic honeycomb lattice, using the linear-spin-wave-theory (LSWT)~\cite{TothLake15, SantosSantosDias18,zhang2020magnonic}. We demonstrate that TOM is highly sensitive to the direction of SW propagation, which can be experimentally detected using the magneto-optical Kerr effect, as illustrated in Fig.~\ref{fig1}. 
Besides, we find that the magnon-related TOM is closely linked to magnonic topology, which can be regulated by parameters such as the DMI and Kitaev interaction.
Moreover, we find that the magnitude and direction of the magnetic field can significantly impact the properties of TOM modulated by magnons, offering a strategy for the experimental verification of magnon-related TOM in real materials.

To explore the properties of TOM in a honeycomb lattice, we first examine the properties of the SSC. As shown in Fig.~\ref{fig2}~(a) and Fig.~S1,
a non-zero SSC can be induced by exciting magnons, which can be understood by considering two types of three-spin clusters based on clockwise chirality: normal triangles (e.g., $\bigtriangleup{135}$) and obtuse triangles (e.g.,   $\bigtriangleup{123}$). 
The TOM is related to the SSC via the relation~\cite{lux2018engineering,grytsiuk2020topological,hanke2017prototypical} $\mathbf{L}^\mathrm{TOM}=\kappa^\mathrm{TO}_{ijk}\sum_{\langle ijk \rangle}  \hat{\mathbf{e}}_{ijk} \chi_{ijk}$,
where $\langle ijk \rangle$ represents the nearby three spins $\mathbf S_i$, $\mathbf S_j$, and $\mathbf S_k$, 
and unit vector $ \hat{\mathbf{e}}_{ijk}$ stands for the normal vector of the plane determined by three spins. The constant $\kappa^{\rm TO}_{ijk}$, known as the topological orbital susceptibility, quantifies the strength of the orbital response of electrons to the SSC~\cite{lux2018engineering,grytsiuk2020topological}.
For simplicity, we set $\kappa^{\rm TO}$ to 2$\mu_{\mathrm B}^{-2}$ for normal triangle and 1$\mu_{\mathrm B}^{-2}$ for obtuse triangle.

The considered here effective spin Hamiltonian of a 2D ferromagnet on a honeycomb lattice is given by:
\begin{equation}
\begin{split}
    H=&-\sum_{\braket{i,j}}J_{ij}  \mathbf{S}_i \cdot \mathbf{S}_j +\sum_{\braket{\braket{ij}}}\mathbf{D}_{{ij}} \cdot (\mathbf{S}_i\times \mathbf{S}_j)\\
     &-K\sum_{\braket{ij}^{\gamma}} S_i^{\gamma}S_j^{\gamma}-\sum_iA(\hat{n}_i\cdot \mathbf S_i)^2\, 
\end{split}
\label{Ham_mag}
\end{equation}
where the strength of the nearest-neighbor
isotropic Heisenberg exchange interaction is given by $J_{ij}$, second-nearest-neighbor DMI is given by $\mathbf{D}_{ij}$, the Kitaev interaction by $K$, and single-ion anisotropy
is given by $A$. 
We set  the spin moment length $S$ to 1.5, 
the easy-axis anisotropy energy along $z$ direction to 0.1 meV, and the second-nearest-neighbor DMI vector is considered to be of the form $\mathbf{D}_{ij}=(0, 0, D)$. 
To ensure that the ground state maintains consistent energy, the exchange interaction $J$ and the Kitaev interaction $K$ are constrained by the equation $3J+K=5.8$ meV~\cite{zhang2021kitaev}.  
The eigenvalues and eigenvectors for each wave vector $\mathbf k$ are determined by the LSWT, and the topological properties of the model are investigated through the magnon Berry curvature $\Omega_{n\mathbf k}^{xy}$~\cite{TothLake15, SantosSantosDias18,zhang2020magnonic,zhang2021kitaev}. The Chern numbers for each magnon branch are obtained by integrating $\Omega_{n\mathbf k}^{xy}$ over the first BZ. As shown in Fig.~\ref{fig2}~(b, c) and Fig. S2, both the Kitaev interaction and the second-nearest-neighbor DMI can significantly influence the Chern numbers and the distribution of
$\Omega_{n\mathbf k}^{xy}$. 
Additional details regarding the model and methods can be found in the supplementary materials~\cite{SupplementalMaterial}.

The magnitude of SSC can be modulated dynamically by spin excitations~\cite{zhang_imprinting_2020}. Following this line of thought, we can imagine that the same mechanism can be used to regulate and imprint topological orbital magnetism by injecting or exciting specific magnon modes, which can be easily realized experimentally, especially for acoustic magnons~\cite{PhysRevLett.123.047204,2023Hybrid,2023Spin}.
Therefore, it is crucial to elucidate the relationship between the SSC and TOM for a specific wave vector $\mathbf k$ of a magnon. Utilizing the orbital electron-magnon coupling theory proposed in our previous work~\cite{zhang_imprinting_2020}, the SSC-dependent TOM can be calculated. 
The local TOM for the $n$th-branch can be expressed as $L_{n\mathbf k}^\mathrm{TOM}=\kappa^\mathrm{TO} \braket{\Psi_{n\mathbf k}|\chi(\mathbf k)|\Psi_{n\mathbf k}}$~\cite{zhang_imprinting_2020}, with $|\Psi_{n\mathbf k}\rangle$  and $\chi(\mathbf k)$ respresenting the wave function and $\mathbf k$-dependent scalar spin chirality, respectively, as shown in Fig.\ref{fig2}.
Notably, the values of the local TOM for optical and acoustic branches are always opposite due to their different "vibrational" character. Furthermore, it is clear that the local TOM is related to the topological properties of the system, as reflected in distinct Chern numbers.
To analyze the magnitude of local TOM for different wave vectors of magnons, the distribution of the local TOM for the acoustic branch in the first Brillouin zone is plotted in Fig.~\ref{fig2}~(d) with parameters $K=5.2$~meV and $D=0$ meV. The results show that the maximum value occurs at the $\mathbf K$ point, while the local TOM remains zero along the $\mathbf{\Gamma}\mathbf M$-path, which can be attributed to the zero SSC there (see Fig.~\ref{fig1} and Fig. S1).

Magnons follow the Bose distribution, and a net magnon-mediated electronic orbital magnetization can be regulated with temperature~\cite{zhang_imprinting_2020, lipeng-2022}. Besides, the magnetization of the system is temperature-dependent, with a magnetic phase transition occurring when the temperature exceeds the Curie-Weiss temperature in ferromagnetic materials. 
To explore temperature-related TOM, the Curie-Weiss temperature
$T_{\mathrm C}$ of the spin system is estimated from the mean-free-theory (MFT), as given by the equation $
k_\mathrm B T_{\mathrm C}=\frac{2S(S+1)}{3}(A+\frac{3}{2}J_1+\frac{1}{2}K)$~\cite{lee2020fundamental}, resulting in a calculated $T_{\mathrm C}$  of about 85~K. 
In our work, TOM is linked to magnons, which are temperature-dependent. To simplify the analysis, we assume the temperature-dependent spin length $S(T)$
that follows the relation: $S(T)= S(1-T/T_{\mathrm C})^{\beta}$, 
with $\beta$ set to 0.3~\cite{Janke-1993,PhysRevB.103.014432}.
Then, the temperature-dependent $\chi(T)$ is expressed as $\chi(T)=\mathbf S_i(T)\cdot(\mathbf S_j(T)\times \mathbf S_k(T))$,  and the local TOM for the $n$-th magnon branch at temperature $T$ is given by the equation:
\begin{equation}
   \ell_n(\mathbf k, T)= \kappa^\mathrm{TO} \braket{\Psi_{n\mathbf k}|\chi(\mathbf k, T)|\Psi_{n\mathbf k}} n_\mathrm{B}(\epsilon_{n\mathbf k}),
    \label{local_TOM}
\end{equation}
where the  $n_\mathrm{B}(\epsilon)=[\exp{(\beta\epsilon)}-1]^{-1}$ is the Bose distribution function with $\beta=1/k_\mathrm{B}T$. The distributions of $\ell_1(\mathbf k, 5)$ and $\ell_1(\mathbf k, 20)$ for the scenario  in Fig.~\ref{fig2}~(d) are shown in Fig.~\ref{fig2}~(e). 
At low temperatures (e.g. 5~K), only magnons with low eigenvalues are excited, and the local TOM contribution is concentrated around the $\mathbf{\Gamma}$ point. As the temperature rises,
magnons with higher eigenvalues become excited, shifting the peak of $\ell_1(\mathbf k, 20)$  to the $\mathbf K/\mathbf K'$ point. 
Moreover,  as shown in Fig.~\ref{fig2}~(c, d, e), $\ell_1(\mathbf k, T)$ remains consistently zero along the $\Gamma-\mathbf K/\mathbf K'$ path, which represents a SW along the armchair direction. 
This spin-wave-regulated TOM behavior in the honeycomb lattice differs significantly from that in other spin lattices, such as the kagome lattice~\cite{zhang_imprinting_2020}, offering a distinctive signature for the presence of TOM. Additional details concerning the wave vector and temperature dependence of TOM are provided in Fig.~S4-5.

The overall TOM per unit cell $\braket{L^\mathrm{TOM}}_\mathrm T$ can be obtained by integrating $\ell_n(\mathbf k, T)$ over the first Brillouin zone across different magnonic branches. As shown in the inset of Fig.~\ref{fig2}(e), the magnitude of TOM initially increases with rising temperature, then gradually decreases, and eventually reaches zero at the Curie temperature. This behavior is attributed to the absence of coherent excitations which can mediate a net chirality as the temperature increases beyond $T_\mathrm C$.
Additionally, $\braket{L^\mathrm{TOM}}_\mathrm T$ as the function of temperature, DMI, and Kitaev interaction are presented in Fig.~\ref{fig2}~(f) and Fig.~S6, separately. These results demonstrate that TOM is influenced by DMI and Kitaev interactions, which govern the system's topological chirality.

\begin{figure}
\centering\vspace{-1pt}
\includegraphics[width=\linewidth]{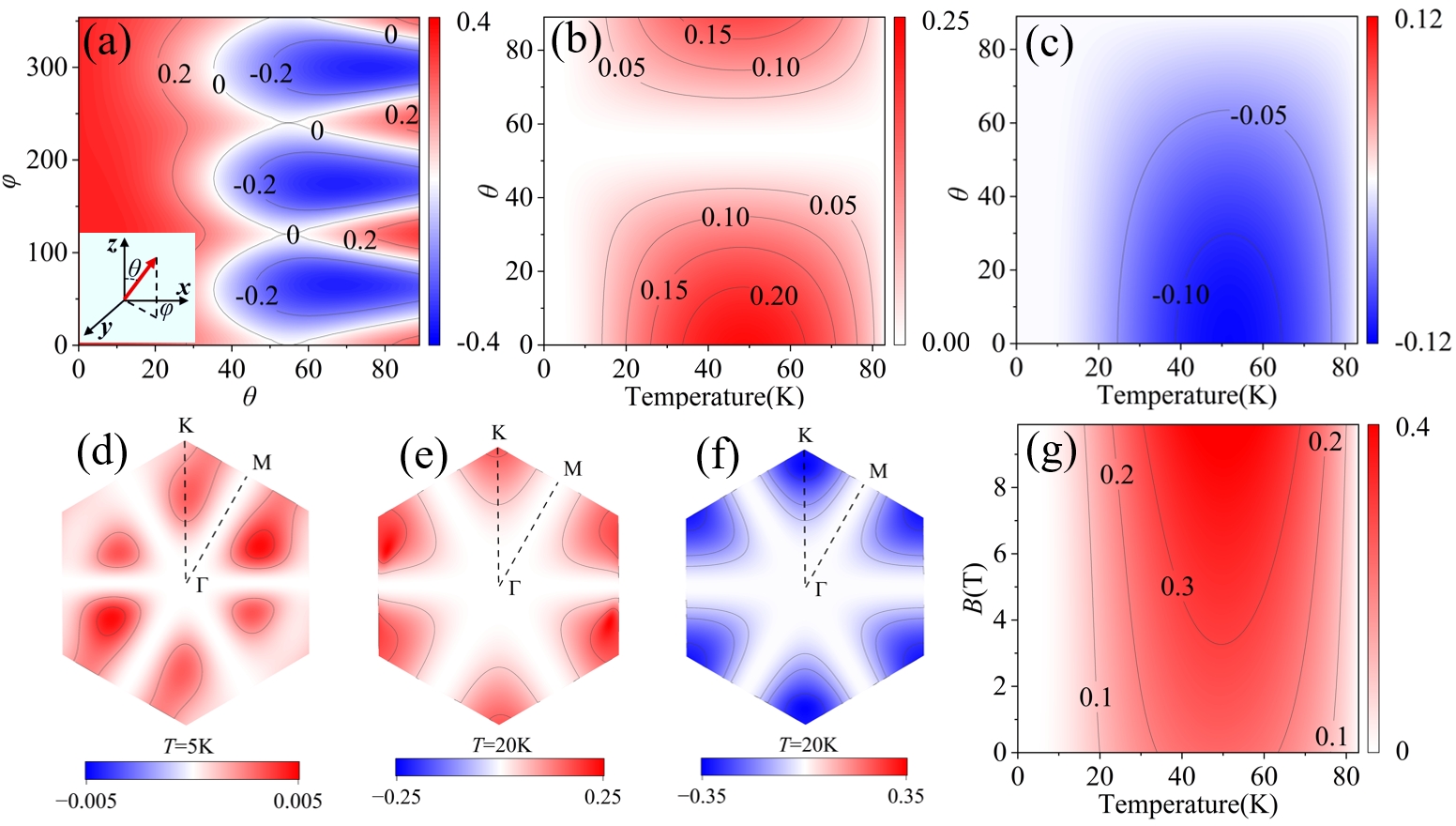}
\caption{ (a)~Evolution of TOM at a temperature of 50~K as a function of polar angle $\theta$ and azimuthal angle $\phi$. The inset represents the direction of the magnetization. (b, c)~ Dependence of the TOM on $\theta$ and temperature.  Both (a, b) refer to the Heisenberg-Kitaev model with $D$=0 meV and $K$=5.2 meV, and (c) corresponds to the Heisenberg-DMI model with $K$=0 meV and $D$=0.3 meV. 
The distribution of $L_{1\mathbf k}^\mathrm{TOM}$ with $\theta=30^\circ$, $\phi=30^\circ$ at 5~K (d) and 20~K (e), and $\theta=90^\circ$, $\phi=60^\circ$ at 20~K (f), separately. (g)~ The TOM as a function of $B$ and $T$ for the Heisenberg-Kitaev model. Here, the direction of the magnetic field is out-of-plane.
The unit of the color map for all figures is $\mu_{\mathrm B}$. }
    \label{fig3}
\end{figure}

In a ferromagnetic kagome system, as we demonstrated previously, TOM can be influenced by a magnetic field~\cite{lipeng-2022, zhang_imprinting_2020}. We now systematically investigate this aspect for a ferromagnetic honeycomb lattice. We begin by examining the effect of magnetization direction on TOM properties, defining the spin direction as ${\mathbf{S}} = \textup{S} (\sin\theta\cos\phi, \sin\theta\sin\phi, \cos\theta)$, where $\theta$ and $\phi$ represent the polar angle and azimuthal angle, respectively.
To facilitate the investigation, we simplified  Eq.~\ref{Ham_mag} into the Heisenberg-Kitaev and the Heisenberg-DMI models, with the corresponding results shown in Fig.~\ref{fig3}~(a,b) and Fig.~\ref{fig3}~(c), separately. 
TOM as a function of $\theta$ and $\phi$  in the Heisenberg-Kitaev model at 50~K is presented in Fig.~\ref{fig3}~(a). 
Both $\theta$ and $\phi$ affect TOM in the Heisenberg-Kitaev model, while TOM is more sensitive to $\theta$ in the Heisenberg-DMI model (as shown in Fig.~\ref{fig3}~(c) and Fig.~S7), consistent with the spin-direction-dependent band gaps and topological properties ~\cite{zhang2021kitaev}.
The distribution of local TOM in the BZ is plotted, and the values of $\ell_1(\mathbf k, 5)$ and $\ell_1(\mathbf k, 20)$ for different $\theta$ and $\phi$ in the Heisenberg-Kitaev model are shown in Fig.~\ref{fig3}~(d-f), respectively. 
The results suggest that the $C_6$ symmetry of the TOM may be broken when $\theta$ and $\phi$ are nonzero, especially at low temperatures. Furthermore, comparing Fig.~\ref{fig3}~(e)  and (f), it is evident that for a given magnon, the generated TOM can be strongly modulated by altering the system's magnetization direction.

The magnetic field not only affects the magnetization direction but also directly couples to both spin and orbital moments. The corresponding Zeeman term in the Hamiltonian is given by $H_{\mathrm B}=
     -\mathbf{B} \cdot \left(\mu_{\mathrm B} \sum_{ijk} \mathbf L^{\mathrm{TOM}}_{ijk}  + \mu_{\mathrm B}g\sum_i \mathbf{S}_i\right)$,
where  {\it g}-factor is chosen as 2. To maintain consistency with the direction of TOM, the magnetic field is defined as $\mathbf B=(0, 0, B)$T in this section.  
The orbital-Zeeman coupling introduces chirality and affects the magnonic topology, which in turn influences the magnitude of TOM.
As shown in Fig.~\ref{fig3}~(g) and Fig.~S8, both topological properties and the magnitude of TOM are modulated by the value of $B$. Therefore, by adjusting the direction and magnitude of the magnetic field, the magnonic topology and TOM can be significantly controlled in a ferromagnetic honeycomb lattice. This modulation may provide strong evidence for the existence of TOM.

\begin{figure}
\centering\vspace{-1pt}
\includegraphics[width=\linewidth]{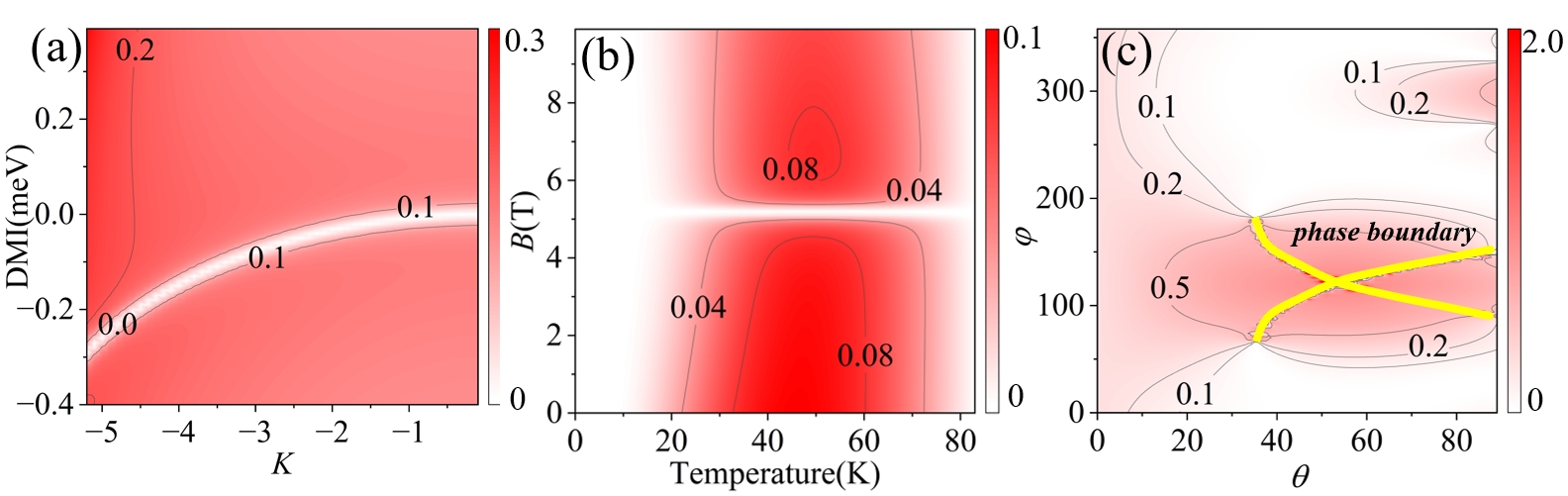}
\caption{ (a)~Dependence of the orbital Nernst conductivity $\kappa_\mathrm{ON}^{xy}$ on the strength of Kitaev interaction and DMI at 50~K. (b)~ The value of $\kappa_\mathrm{ON}^{xy}$ modulated by the out-of-plane magnetic field for the Heisenberg-DMI model with $D$=$-$0.3 meV, and $K$=0 meV.
(c)~Evolution of $\kappa_\mathrm{ON}^{xy}$ at 50~K as a function of polar angle $\theta$ and azimuthal angle $\phi$ for the Heisenberg-Kitaev model. 
The unit of all the color maps is $k_B/2\pi$.}
\label{fig4}
\end{figure}

A finite SSC can be triggered by thermally activated magnons~\cite{go2024scalar}, and the TOM current can be produced by the spatial temperature gradients, a phenomenon known as the topological orbital Nernst effect (TONE)~\cite{zhang_imprinting_2020}. In this section, the  topological orbital Nernst conductivity (TONC) in ferromagnetic honeycomb lattice is calculated following:
\begin{equation}
\begin{split}
    \kappa_\mathrm{ON}^{xy}=-\frac{k_\mathrm{B}}{4\pi^2\mu_\mathrm{B} }\sum_n\int\displaylimits_{\rm BZ} c_1(n_\mathrm{B}(\epsilon_{n\mathbf k}))\,\Omega_{n\mathbf k}^{xy}\ell_n(\mathbf k, T) \, d\mathbf k,
\end{split}
\label{eq:tone}
\end{equation}
with 
$c_1(\tau)=\int_0^{\tau}\ln[(1+t)/t]dt=(1+\tau)\ln(1+\tau)-\tau\ln \tau$. 
According to Eq.~\ref{eq:tone}, $\kappa_\mathrm{ON}^{xy}$  depends on $\Omega_{n\mathbf k}^{xy}$ and $\ell_n(\mathbf k, T)$, which 
are influenced by the DMI, Kitaev interaction, and magnetic field. 
We examine how these parameters regulate
$\kappa_\mathrm{ON}^{xy}$.  The magnitude of $\kappa_\mathrm{ON}^{xy}$ at 50~K, influenced by DMI and Kitaev interaction, is shown in Fig.~\ref{fig4}~(a). It demonstrates that $\kappa_\mathrm{ON}^{xy}$ is divided into two regions by the topological phase boundary, with the maximum value occurring when both $K$ and $D$ are large.
Besides, introducing a magnetic field can strongly modulate $\kappa_\mathrm{ON}^{xy}$. For instance, $\kappa_\mathrm{ON}^{xy}$ can be controlled by
an out-of-plane magnetic field (Fig.~\ref{fig4}~(b)) and by the magnetization direction, determined by values of $\theta$ and $\phi$ (Fig.~\ref{fig4}~(c)).
Moreover, a comparison is made between the magnon spin Nernst effect~\cite{matsumoto2011theoretical,cheng2016spin,kovalev2016spin} and TONE. As shown in Fig.~\ref{fig4} (a) and Fig.~S9, the magnitude of TONC is comparable to the magnon spin Nernst conductivity (MSNC). Besides, the direction of TONC remains unchanged during a topological phase transition, which is different from MSNC.
It is worth noting that the TONC is closely related to $\kappa^{\rm TO}$, and the estimated conductivity may be lower when a range of material-dependent values of $\kappa^{\rm TO}$ is considered~\cite{hoffmann2015topological,hanke2016role,hanke2017prototypical,grytsiuk2020topological}. A more detailed comparison of the temperature dependence of $\kappa_\mathrm{ON}^{xy}$ and $\kappa_\mathrm{N}^{xy}$ 
is shown in Fig.~S10 and Fig.~S11.

In this work, we combine orbitronics and magnonics on a ferromagnetic honeycomb lattice through our proposed model of magnon-mediated TOM. We find that various spin interactions, such as the Kitaev interaction and DMI, not only induce and control TOM but also affect the spin and TOM transport properties via magnon excitations, which imprint their topology in the spin and orbital response. Since the Kitaev interaction and DMI can be easily modulated, for instance, through external strain~\cite{xu2020possible} or electric field~\cite{dai2023electric}, we believe that TOM characteristics can be efficiently tuned. Moreover, we show that the accumulation of electronic orbital moments can be controlled by different SW vectors, which can be detected via the magneto-optical Kerr effect~\cite{OHE2023,shinagawa2000faraday} or scanning transmission electron microscopy~\cite{idrobo2024direct}, among other techniques. 

It is important to highlight that the orbital magnetism discussed in our work is distinct from the orbital moments of magnons previously investigated in antiferromagnetic honeycomb~\cite{go2024magnon} and kagome lattice~\cite{PhysRevLett.125.117209}. Unlike the orbital magnetic moment of magnons, TOM is significantly easier to detect~\cite{hanke2017prototypical,lux2018engineering,grytsiuk2020topological}, with experimental evidence of magnon-mediated TOM already reported in Cu(1,3-bdc)~\cite{lipeng-2022}. Additionally, magnetic fields can be used to experimentally confirm the existence of TOM in real systems. On one hand, TOM is influenced by the system's chirality and, in turn, can modulate the shape of magnon dispersion and the topology of magnonic states through coupling with external magnetic fields. On the other hand, we discovered that the direction of magnetization can strongly modulate the magnitude of TOM. 

In summary, our work offers significant insights into the interplay between spin excitations and orbital magnetism, demonstrating that electron-magnon orbital coupling creates new opportunities for integrating spin-orbitronics into magnonic devices. Furthermore, the exploration of spin-wave-regulated TOM introduces a new, so far not considered variable into the physics of hybrid magnon-based states driven by magnon-phonon coupling~\cite{kunz2024coherent} and magnon-photon coupling~\cite{hirosawa2022magnetoelectric}. This may prove crucial in studying the emergence, dynamics, and properties of novel quasiparticles.

\section{Acknowledgements}\label{Sec:Acknowledgements}
This project was supported by the  National Natural Science Foundation of China (1240041438, 12347156), the Natural Science Foundation of Jiangsu Province (No. BK20230516), China Postdoctoral Science Foundation (2024M761185), the Scientific Research Startup of Jiangsu University (No. 5501710001). Lichuan Zhang acknowledges the funding of the overseas postdoctoral talent attraction program from China. We also gratefully acknowledge financial support by the Deutsche Forschungsgemeinschaft (DFG, German Research Foundation) $-$ TRR 288 $-$ 422213477 (project B06), TRR 173/2 $-$ 268565370 (project A11).  This work was supported by the EIC Pathfinder OPEN grant 101129641 "OBELIX".



\bibliography{Kitaev}

\end{document}